\documentclass[letter,aps,twocolumn]{revtex4-1}
\usepackage{graphics}
\usepackage{graphicx}
\usepackage{amssymb}
\usepackage{epsfig}
\usepackage{color}
\usepackage{ulem}
\newlength{\upit}\upit=0.1truein

\newcommand{\ltappr}{{{\lower4pt\hbox{$<$} } \atop \widetilde{ \ \ \ }}}
\newlength{\bxwidth}\bxwidth=1.5 truein

\newlength{\figwidth}
\newlength{\shift}
\newcommand \bea {\begin{eqnarray} }
\newcommand \eea {\end{eqnarray}}

\begin{document}

\title{Kondo insulator to semimetal transformation tuned by spin-orbit coupling}

\author{S. Dzsaber$^{1}$, L. Prochaska$^{1}$, A. Sidorenko$^{1}$,
G.~Eguchi$^{1}$, R.~Svagera$^{1}$, M.~Waas$^{1}$, A. Prokofiev$^{1}$, Q.
Si$^{2}$, and S. Paschen$^{1,2,\ast}$}

\affiliation{$^1$Institute of Solid State Physics, Vienna University of
Technology, Wiedner Hauptstr.\ 8-10, 1040 Vienna, Austria}

\affiliation{$^2$Department of Physics and Astronomy, Rice University, Houston, Texas 77005, USA}


\date{\today}

\begin{abstract}
Recent theoretical studies of topologically nontrivial electronic states in
Kondo insulators have pointed to the importance of spin-orbit coupling (SOC) for
stabilizing these states. However, systematic experimental studies that tune the
SOC parameter $\lambda_{\rm{SOC}}$ in Kondo insulators remain elusive. The main
reason is that variations of (chemical) pressure or doping strongly influence
the Kondo coupling $J_{\text{K}}$ and the chemical potential $\mu$ -- both
essential parameters determining the ground state of the material -- and thus
possible $\lambda_{\rm{SOC}}$ tuning effects have remained unnoticed. Here we
present the successful growth of the substitution series
Ce$_3$Bi$_4$(Pt$_{1-x}$Pd$_x$)$_3$ ($0 \le x \le 1$) of the archetypal
(noncentrosymmetric) Kondo insulator Ce$_3$Bi$_4$Pt$_3$. The Pt-Pd substitution
is isostructural, isoelectronic, and isosize, and therefore likely to leave
$J_{\text{K}}$ and $\mu$ essentially unchanged. By contrast, the large mass
difference between the $5d$ element Pt and the $4d$ element Pd leads to a large
difference in $\lambda_{\rm{SOC}}$, which thus is the dominating tuning
parameter in the series. Surprisingly, with increasing $x$ (decreasing
$\lambda_{\rm{SOC}}$), we observe a Kondo insulator to semimetal transition,
demonstrating an unprecedented drastic influence of the SOC. The fully
substituted end compound Ce$_3$Bi$_4$Pd$_3$ shows thermodynamic signatures of a
recently predicted Weyl-Kondo semimetal.
\end{abstract}


\maketitle

Topological phases \cite{NatPhys16.1} in condensed matter systems
\cite{Has10.1,Qi11.1}, most recently encompassing also gapless variants
\cite{NatMater16.1}, continue to attract great attention. To date, work has
mostly focused on weakly correlated materials, but it is clear that yet more
exotic physics may be discovered in strongly correlated settings \cite{Sch16.2}.
Thus, the proposal that Kondo insulators \cite{Aep92.1,Ris00.1} -- some of the
most strongly correlated materials -- exhibit topologically nontrivial metallic
surface states \cite{Dze10.1} was taken up enthusiastically and triggered many
experimental studies, most notably on SmB$_6$ \cite{Dze16.1}. A key ingredient
for a topologically nontrivial electronic structure in Kondo insulators is the
strong spin-orbit coupling (SOC) of the heavy lanthanide $4f$ elements
\cite{Dze10.1}, but the importance of SOC of the conduction electrons that
hybridize with the $4f$ electrons has also been demonstrated \cite{Fen13.1}. In
studies of the periodic Anderson model, the latter was shown to tune between
different phases, including topological and topologically trivial Kondo
insulators \cite{Fen13.1}, Dirac-Kondo semimetals \cite{Fen16.1}, and most
recently Weyl-Kondo semimetals \cite{Lai16.1}. To link such studies directly to
experiment it would be highly desirable to find an experimental ``tuning knob''
for SOC in Kondo systems. The availability of parameters that tune the Kondo
interaction strength has been vital to the field of heavy fermion quantum
criticality \cite{Loe07.1,Si13.1}. Here we demonstrate for the first time SOC
tuning in a Kondo insulator.

In Kondo insulators the Kondo interaction between localized ($4f$ and less
frequently $5f$ or $3d$) and itinerant (typically $d$) electrons opens a narrow
gap -- of the order of 10\,meV -- in the electronic density of states at the
Fermi level \cite{Aep92.1,Ris00.1}. Among the archetypal cubic Kondo insulators
that have been studied for decades \cite{Aep92.1,Ris00.1}, Ce$_3$Bi$_4$Pt$_3$
appears as an ideal starting material for our study. As a Ce-based system it
represents the conceptually simple situation of a single $4f$ electron as the
localized species. Its Kondo insulating ground state is well established
\cite{Hun90.1,Sev91.1,Buc94.1} and quite robust: The Kondo insulator gap
persists up to 145\,kbar \cite{Coo97.1} and is closed only for magnetic fields
as high as 40\,T \cite{Jai00.1}. Ce$_3$Bi$_4$(Pt,Pd)$_3$ crystallizes in a cubic
structure of space group $I\bar43d$ with a noncentrosymmetric unit cell
containing 40 atoms \cite{Dzs17.1SI}. All three constituting
elements occupy unique crystallographic sites and form three sublattices
that all lack inversion symmetry. Interestingly, mirroring Ce at the unit cell
center transforms it into Pt (Pd) and vice versa (left inset in
Fig.\,\ref{Fig:Temp}), reflecting the direct involvement of these elements in
defining the noncentrosymmetric structure. Theoretical studies have suggested
that Ce$_3$Bi$_4$Pt$_3$ hosts topologically nontrivial states
\cite{Dze10.1,Cha16.1}, but these could so far not be experimentally resolved
\cite{Wak16.1}.

Generally, chemical substitution with atoms of sizeable mass difference seems a
promising route for SOC tuning because the atomic number $Z$ (or the mass)
enters the SOC parameter as $\lambda_{\rm{SOC}}\sim Z^4$ \cite{Sha14.1}. In
Kondo insulators, however, a clever choice of the type of substitution has to be
made. A substitution of the $4f$ element (Ce) breaks the translational symmetry
of the local moment sublattice, which leads to a loss of Kondo coherence. To
keep the Kondo lattice intact, substitutions should therefore be limited to the
nonmagnetic elements (Bi and Pt). Most relevant for the Kondo interaction are
the transition metal $d$ electrons. Indeed, photoemission experiments evidence
the presence of Pt $5d$ states near the Fermi level \cite{Tak99.1}, suggesting
that a substitution of Pt by another transition element would be most relevant.
This is further underpinned by the fact that the Ce atoms in
Ce$_3$Bi$_4$(Pt,Pd)$_3$ have only Pt (Pd) as nearest neighbors (upper right
inset in Fig.\,\ref{Fig:Temp}). The second constraint is that Kondo insulators,
just as heavy fermion metals, react sensitively to even small changes of
chemical pressure. Thus, to keep the Kondo coupling $J_{\text{K}}$ tuning
minimal, isosize substitutions should be used. Finally, Kondo insulators being
insulators naturally makes them react strongly to changes in electron count and
thus in the chemical potential $\mu$, which favors isoelectronic substitutions
(without carrier ``doping'').

Previous studies have failed to separate these different effects. For instance,
in the Kondo semimetal CeNiSn, isoelectronic but non-isosize substitutions of Ni
by Pt or Pd close the Kondo (pseudo)gap as a consequence of the increased unit
cell volume and hence the decreased $J_{\text{K}}$ \cite{Nis93.1,Kas88.1}. In
Ce$_3$Sb$_4$Pt$_3$, the non-isosize and non-isoelectronic substitutions of Pt by
Cu and Au both suppress the Kondo insulating state although Cu doping results in
a reduced \cite{Jon99.1} and Au doping in an increased unit cell volume
\cite{Kat96.1}. Thus, here the change in $\mu$ dominates.

\begin{figure}[t!]
\begin{center}
\includegraphics[width=0.47\textwidth]{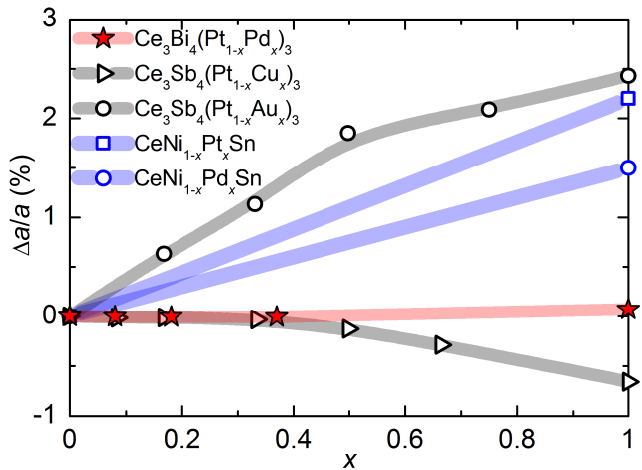}
\vspace{-0.2cm}

\caption{(Color online) Relative change of lattice parameter $a$ as function of
the Pd content $x$ in Ce$_3$Bi$_4$(Pt$_{1-x}$Pd$_x$)$_3$ (red), on a scale
relevant to other archetypal Kondo systems
\cite{Kat96.1,Jon99.1,Nis93.1,Kas88.1} (open symbols). Lines are guides to the
eyes.}
\label{Fig:Lattice}
\end{center}
\vspace{-0.9cm}
\end{figure}

Surprisingly, no substitution series of Ce$_3$Bi$_4$Pt$_3$ other than Ce-La
replacements \cite{Sch92.1,Pie08.1} have yet been studied. Here we present first
results on the series Ce$_3$Bi$_4$(Pt$_{1-x}$Pd$_x$)$_3$ and show that it
ideally qualifies to study pure $\lambda_{\rm{SOC}}$ tuning. The $4d$ transition
metal Pd is much lighter than the $5d$ transition metal Pt (atomic weight 106.42
instead of 195.084) and thus an increase of the Pd content $x$ should sizeably
reduce the conduction electron $\lambda_{\rm{SOC}}$. By contrast, as Pt and Pd
are isoelectronic, there is minimal $\mu$ tuning. Furthermore, as will be show
below, there is minimal $J_{\text{K}}$ tuning.

Single crystals of Ce$_3$Bi$_4$(Pt$_{1-x}$Pd$_x$)$_3$ were grown using a
modified Bi flux method \cite{Dzs17.1SI}. The substitution levels $x$
were determined by EDX measurements \cite{Dzs17.1SI}. The lattice
parameter $a$ across the sample series is shown in Fig.\,\ref{Fig:Lattice}. The
accumulated relative change $\Delta a/a(x) = [V(x)/V(x=0)]^{1/3}-1$, where
$V(x)$ denotes the unit cell volume for a given substitution level $x$, is only
0.069\% at $x = 1$. This is extremely small compared to substitutions in related
materials (Fig.\,\ref{Fig:Lattice}). Thus, unlike in these other substitution
series, $J_{\rm{K}}$ tuning by chemical pressure can be excluded as dominating
factor in Ce$_3$Bi$_4$(Pt$_{1-x}$Pd$_x$)$_3$. 

\begin{figure}[t!]
\begin{center}
\includegraphics[width=0.47\textwidth]{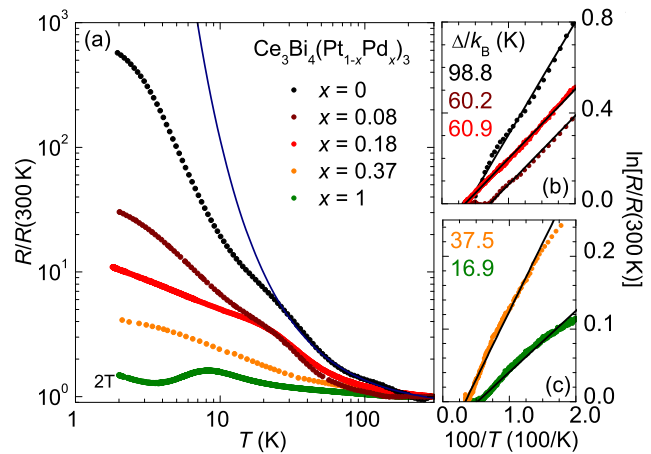}
\vspace{-0.2cm}

\caption{(Color online) (a) Temperature-dependent electrical resistance $R(T)$,
normalized to its room temperature value $R(300$\,K), on a $\log-\log$ scale for
all investigated Ce$_3$Bi$_4$(Pt$_{1-x}$Pd$_x$)$_3$ samples. An exponentially
activated resistivity, fitted to the high-temperature regime of the $x=0$
sample, is shown as blue line. (b,\,c) Arrhenius plots for data above 50\,K. The
solid lines are the fits (see text). The data for the $x = 1$ sample were taken
at 2\,T to suppress a spurious superconducting resistance drop at 3\,K that has
no correspondence in magnetization and specific heat measurements. This magnetic field has no effect on the resistance data above 3\,K.}
\label{Fig:Resistivity}
\end{center}
\vspace{-0.7cm}
\end{figure}

\begin{figure}[t!]
\begin{center}
\includegraphics[width=0.47\textwidth]{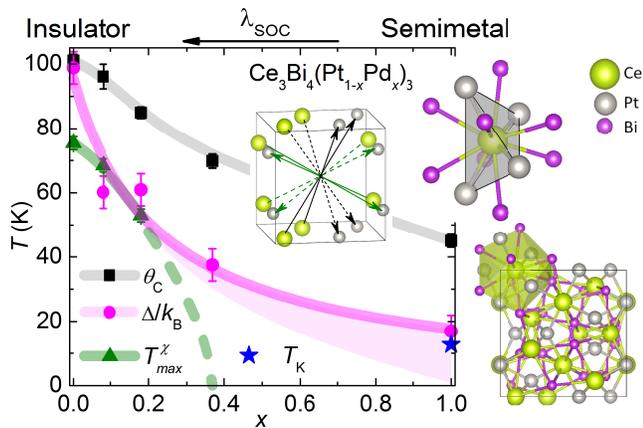}
\vspace{-0.2cm}

\caption{(Color online) Characteristic temperature scales for
Ce$_3$Bi$_4$(Pt$_{1-x}$Pd$_x$)$_3$ as function of $x$. $T^{\chi}_{\rm{max}}$ and
$\Theta_{\rm{C}}$ are taken from Fig.\,\ref{Fig:SQUID}(a) and the Curie-Weiss
fits in Fig.\,\ref{Fig:SQUID}(b), respectively. $\Delta/k_{\rm{B}}$ is taken
from the Arrhenius fits in Fig.\,\ref{Fig:Resistivity}(b,c). The Kondo temperature $T_{\rm K}$ is taken from the entropy analysis in Fig.\,\ref{Fig:Cp}(c). The full lines are guides to the eyes. The shaded pink area represents the fact that, for $x=1$, the Arrhenius fit loses significance and $R(T)$ becomes compatible with a gapless state (see text). The dashed green line indicates that the maximum in $\chi(T)$ is suppressed to below 2\,K for $x=0.37$. The arrow on the top axis indicates that an {\it{in}}crease of the conduction-electron $\lambda_{\rm{SOC}}$ (corresponding to a {\it{de}}crease of $x$) drives the system from a semimetallic to an insulating state. The structure sketches illustrate the unit cell (bottom), the environment of Ce with 4 nearest Pt/Pd (3.01\,\AA) and 8 next-nearest Bi (3.41\,\AA) neighbors (top), and the lack of a center of inversion for (selected) Ce and Pt atoms.}
\label{Fig:Temp}
\end{center}
\vspace{-0.9cm}
\end{figure}

Figure\,\ref{Fig:Resistivity} shows the temperature dependence of the electrical
resistance $R(T)$ of all investigated Ce$_3$Bi$_4$(Pt$_{1-x}$Pd$_x$)$_3$
crystals, normalized to the respective room temperature value (180 to
350\,$\mu\Omega$cm, in good agreement with the published value of
220\,$\mu\Omega$cm for Ce$_3$Bi$_4$Pt$_3$ \cite{Hun90.1}; small differences are
attributed to the poorly defined geometrical factors of the small single
crystals).  With increasing $x$, $R$ at low temperatures is successively
reduced, corresponding to a gradual closing of the Kondo insulator gap. This can
be quantified by Arrhenius fits $R = R_0 \exp[\Delta/(2k_{\rm{B}}T)]$ to the
high-temperature data [Fig.\,\ref{Fig:Resistivity}\,(b,c)], where $\Delta$ is
the gap width and $k_{\rm{B}}$ is the Boltzmann constant. The continuous
decrease of $\Delta$ with $x$ is shown in Fig.\,\ref{Fig:Temp}. It is remarkable
that the (full) substitution of Bi by the much lighter isoelectronic element Sb
has an entirely different effect: Instead of closing the Kondo insulator gap it
strongly enhances it to 1080\,K \cite{Kat96.1}. This must be due to the smaller
size of Sb, that leads to a lattice parameter reduction by 2.3\% \cite{Kat96.1},
and thus a stronger hybridization, similar to the gap opening under hydrostatic
pressure \cite{Coo97.1}.

\begin{figure}[t!]
\begin{center}
\includegraphics[width=0.47\textwidth]{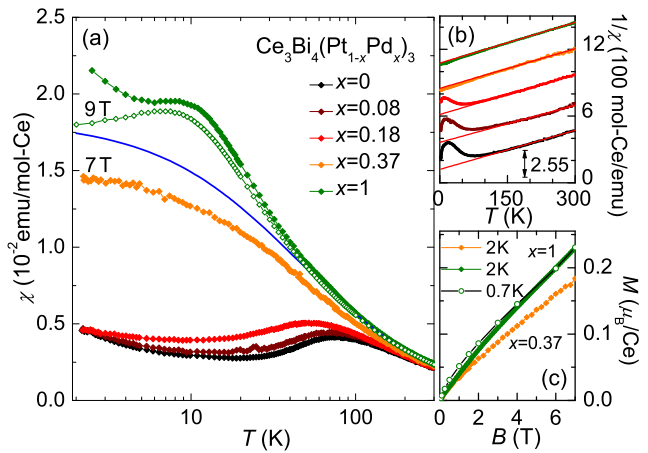}
\vspace{-0.2cm}

\caption{(Color online) (a) Temperature-dependent magnetic susceptibility
$\chi(T)$ of all investigated Ce$_3$Bi$_4$(Pt$_{1-x}$Pd$_x$)$_3$ samples,
together with the Curie-Weiss fit for the $x = 1$ data (blue line, from b). (b)
$\chi^{-1}$ vs $T$ along with the Curie-Weiss fits that determine the
paramagnetic Weiss temperature $\Theta_{\rm{C}}$. The data for $x > 0$ are
successively shifted by 255 mol-Ce\,Oe/emu for clarity. Unless specified, the
field for the data in (a) and (b) was 100\,mT. (c) Selected magnetization vs
field $M(B)$ isotherms. The slight nonlinearities at low fields indicate that
the low-temperature upturn of $\chi(T)$, seen most clearly for the $x=1$ sample
in (a), is a readily saturable by small fields.}
\label{Fig:SQUID}
\end{center}
\vspace{-0.9cm}
\end{figure}

\begin{figure}[t]
\begin{center}
\includegraphics[width=0.47\textwidth]{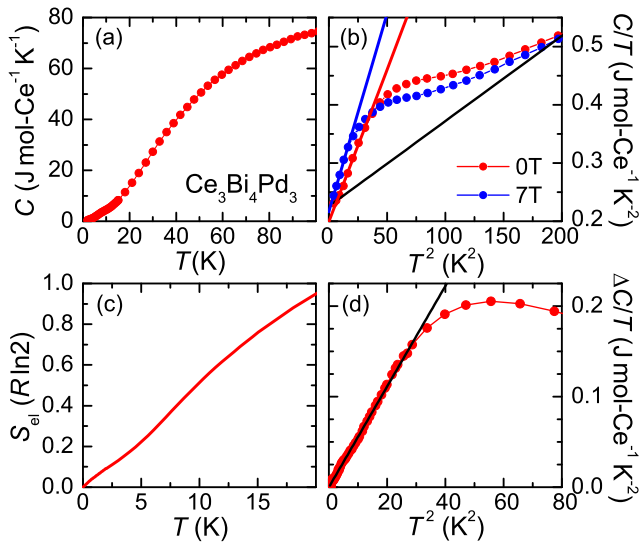}
\vspace{-0.2cm}

\caption{(Color online) (a) Temperature-dependent specific heat $C(T)$ of
Ce$_3$Bi$_4$Pd$_3$ below 100\,K. (b) $C/T$ vs $T^2$ below 14\,K, in zero
field and 7\,T. The slope $\beta$ of the black straight line corresponds to the
phonon contribution $C_{\rm{ph}}$; the red and blue lines with steeper slope
$\beta^{\prime}$ are interpreted as Weyl contributions (see text). (c) Entropy
of electronic specific heat $C_{\rm{el}} = C - C_{\rm{ph}}$ vs temperature. (d)
With the lowest-$T$ CDW and NFL contributions (Supplemental Fig.\,S3)
subtracted, linear-in-$T^2$ behavior (black line) of $C_{\rm{el}}/T$ prevails
from 0.4 to 5.5\,K.}
\label{Fig:Cp}
\end{center}
\vspace{-0.8cm}
\end{figure}

Gap values that are much smaller than the lower boundary of the fitting range
are unphysical. This is the case for Ce$_3$Bi$_4$Pd$_3$. Its energy gap of
16.9\,K should thus be taken with caution -- and rather as indication for the
absence of a well-defined gap (as indicated by the pink shaded region in
Fig.\,\ref{Fig:Temp}). In fact, $R$ of this sample depends only very weakly on
temperature, which is characteristic of semimetals. At 8.3$\,$K, a broad local
maximum is observed. As will be shown below, this feature is echoed by features
in the magnetization (Fig.\,\ref{Fig:SQUID}) and specific heat
(Fig.\,\ref{Fig:Cp}), and is likely due to Kondo interaction in a semimetal.

The magnetic susceptibility of all investigated
Ce$_3$Bi$_4$(Pt$_{1-x}$Pd$_x$)$_3$ samples [Fig.\,\ref{Fig:SQUID}(a)] is well
described by a Curie-Weiss law at high temperatures [Fig.\,\ref{Fig:SQUID}(b)].
The effective magnetic moments, obtained by Curie-Weiss fits between 150 and
300\,K, all agree within error bar with the value of 2.54\,$\mu_{\text{B}}$
expected for a free Ce$^{3+}$ ion. The (negative) paramagnetic Weiss temperature
$\Theta_{\rm{C}}$, which is a measure of the strength of antiferromagnetic (AFM)
correlations driven by the Kondo interaction, decreases continuously in absolute
value (Fig.\,\ref{Fig:Temp}), but remains sizeable even for the $x=1$ sample.
This is in sharp contrast to the Au-Pt and Cu-Pt substitutions in
Ce$_3$Sb$_4$Pt$_3$ which suppress $-\Theta_{\rm{C}}$ from 647\,K in
Ce$_3$Sb$_4$Pt$_3$ to less than 3\,K and 1\,K in Ce$_3$Sb$_4$Au$_3$
\cite{Kat96.1} and Ce$_3$Sb$_4$Cu$_3$ \cite{Jon99.1}, respectively.

A maximum in the magnetic susceptibility, as observed at
$T^{\chi}_{\rm{max}}=75\,$K for the $x=0$ sample [Fig.\,\ref{Fig:SQUID}(a)], is
in good agreement with previous findings \cite{Hun90.1,Buc94.1}. It signals the
onset of Kondo screening associated with the opening of the Kondo insulator gap
\cite{Ris00.1}. With increasing $x$, $T^{\chi}_{\rm{max}}$ is successively
suppressed (Fig.\,\ref{Fig:Temp}). The $x = 0.37$ sample shows no maximum down
to at least 2\,K. Sample-dependent upturns of the susceptibility at the lowest
temperatures, that have no correspondence in neutron scattering experiments
\cite{Sev91.1}, are generally attributed to a Curie tail due to a small amount
of magnetic impurities. 

The magnetic susceptibility of Ce$_3$Bi$_4$Pd$_3$ is qualitatively different.
Below 50\,K, it deviates to values larger than the Curie-Weiss law [blue line in
Fig.\,\ref{Fig:SQUID}(a)] and tends to saturate below 10\,K, characteristics of
heavy fermion metals. The upturn at the lowest temperatures is again suppressed
by magnetic fields.

Further information on the nature of the ground state of Ce$_3$Bi$_4$Pd$_3$ can
be extracted from specific heat $C(T)$ measurements [Fig.\,\ref{Fig:Cp}(a)]. To
determine the electronic specific heat $C_{\rm{el}}=C-C_{\rm{ph}}$ of a heavy
fermion material it is common practice to use the phonon specific heat
$C_{\rm{ph}}$ as determined from its non-$f$ reference material, which is
usually well described by $C/T = \gamma + \beta T^2$. The Sommerfeld coefficient
$\gamma$ represents the electronic contribution and the $\beta$ term the Debye
approximation of the lattice contribution. Indeed, this relation holds for
La$_3$Bi$_4$Pt$_3$, with $\beta = 1.46$\,mJ/(mol-La\,K$^4$) \cite{Hun90.1}. The
specific heat of Ce$_3$Bi$_4$Pd$_3$ displays a pronounced anomaly with respect
to this Debye behavior, which is only slightly shifted to lower temperatures in
a magnetic field of 7\,T [Fig.\,\ref{Fig:Cp}(b)]. On the low-temperature side of
the anomaly, $C/T$ is linear in $T^2$, with a slope $\beta^{\prime}$ that is
sizeably larger than that of the phonon contribution. Similar behavior was seen
in the cubic heavy fermion antiferromagnet CeIn$_3$ below the N\'eel temperature
\cite{Ber79.1,Cor01.1} and attributed to 3D AFM magnons. However, in
Ce$_3$Bi$_4$Pd$_3$, there is no obvious sign of a magnetic phase transition as
the observed anomalies appear too broad to represent such transitions. In
addition, in view of the linear coupling of a magnetic field to a symmetry
breaking order parameter, a stronger suppression would be expected if the
anomaly indeed represented AFM ordering. Finally, the magnetic field enhances
$\beta^{\prime}$, but would be expected to reduce it in a 3D AFM magnon scenario
\cite{Deo73.1,Map06.1}.

Instead, we suggest that the $\beta^{\prime}T^2$ contribution originates from
bulk electronic states with linear dispersion $\varepsilon_{\bf k}=\hbar v^* k$,
with the quasiparticle velocity $v^*$, recently predicted for a Weyl-Kondo
semimetal \cite{Lai16.1}. Such states contribute a volume specific heat of
$7\pi^2/30\cdot k_B[(k_BT)/(\hbar v^*)]^3$ (Supplemental Material of
\cite{Lai16.1}). From our experiments we determine $v^* = 885$\,m/s which is
three orders of magnitude smaller than the Fermi velocity of a simple metal.
Interestingly, the (single ion) Kondo temperature $T_{\rm{K}}$ that we estimate
as the temperature where the electronic entropy [Fig.\,\ref{Fig:Cp}(c)] reaches
$0.65\ln{2}$ per Ce \cite{Des82.1}, is 13\,K ($\sim$\,meV) and thus three orders
of magnitude smaller than the Fermi temperature of a simple metal ($\sim$\,eV).
This further supports the 1000-fold renormalization of the quasiparticle
velocity discussed above.

The lowest-temperature $C_{\rm{el}}/T$ data, plotted on a logarithmic
temperature scale (Supplemental Fig.\,S3), reveal further signs of strong
correlations: A Schottky-like anomaly and a non-Fermi liquid (NFL)-like
$\ln{(1/T)}$ upturn are discerned below 2 and 0.8\,K, respectively. The former
is likely a precursor of the formation of a charge density wave (CDW), a
particle-hole instability that has recently been predicted for Weyl semimetals
with long-range repulsive Coulomb interactions \cite{Wei14.2}. The latter may
indicate that Ce$_3$Bi$_4$Pd$_3$ is close to a quantum critical point, with its
Fermi level slightly away from the Weyl nodes (see \cite{Dzs17.1SI} for
further details). These observations are exciting on their own and clearly call
for further studies. If we model these low-temperature contributions [orange
line in Fig.\,S3] and subtract them from the $C_{\rm{el}}/T$ data, the
$\beta^{\prime}T^2$ term that evidences the Weyl-Kondo semimetal dispersion is
seen down to the lowest temperatures [Fig.\,\ref{Fig:Cp}(d) and Supplemental
Fig.\,S3], thus over more than a decade in temperature.

Figure\,\ref{Fig:Temp} summarizes the effects of increasing Pd content $x$
(bottom axis) and thus decreasing $\lambda_{\rm{SOC}}$ (see top axis): The
absolute value of the paramagnetic Weiss temperature $\Theta_{\rm{C}}$ decreases
with increasing $x$, but remains sizeable even for $x=1$. The Kondo insulator
gap $\Delta$ is reduced with increasing $x$ and cannot be clearly discerned
beyond $x=0.37$. The spin screening temperature $T^{\chi}_{\rm{max}}$ is
likewise reduced with $x$, to below 2\,K for $x=0.37$. For the semimetal
Ce$_3$Bi$_4$Pd$_3$ at $x=1$, it reappears in the form of a Kondo temperature of
13\,K, as determined from the electronic entropy. This evolution, together with
the above discussed thermodynamic features of Ce$_3$Bi$_4$Pd$_3$, strongly
suggests that we have tuned Ce$_3$Bi$_4$(Pt$_{\rm{1-x}}$Pd$_{\rm{x}}$)$_3$ from
a Kondo insulator phase for $x=0$ to the recently predicted Weyl-Kondo semimetal
phase \cite{Lai16.1} at $x=1$. In fact, a topological Kondo insulator to
Dirac-Kondo semimetal transition has recently been shown in an Anderson lattice
model upon reducing $\lambda_{\rm{SOC}}$ \cite{Fen16.1}. For the case of a
noncentrosymmetric crystal structure, a transition between a topological Kondo
insulator and a Weyl-Kondo semimetal \cite{Lai16.1} may thus be theoretically
expected. This exciting perspective calls for additional experiments, to further
probe the topological bulk and surface states predicted for a Weyl-Kondo
semimetal \cite{Lai16.1}, as well as further theoretical studies on its
evolution upon $\lambda_{\rm{SOC}}$ tuning.

Similar SOC tuning studies may also shed light on the topological nature of
other strongly correlated semimetals, such as CeRu$_4$Sn$_6$ \cite{Gur13.1} and
CeNiSn \cite{Sto16.1}, that have been suggested to host topological bulk and/or
surface states \cite{Sun15.2,Cha16.1,Xu16.2}. In the substitution series
Yb$_3$(Rh$_{1-x}$Ir$_x$)$_4$Ge$_{13}$ a metal to AFM insulator crossover was
observed with increasing $x$ \cite{Rai16.1}. Interestingly, the Rh-Ir
substitution is isoelectronic, isostructural, and almost isosize, suggestive of
predominant SOC tuning, though this was not acknowledged in that work.

In conclusion we have presented noncentrosymmetric
Ce$_3$Bi$_4$(Pt$_{1-x}$Pd$_x$)$_3$ as a model system for spin-orbit coupling
(SOC) tuning of a Kondo insulator. A continuous decrease of the SOC strength is
achieved by the isoelectronic, isostructural, and isosize substitution of the
heavy $5d$ element Pt by the much lighter $4d$ element Pd on the nearest
neighbor site of Ce in the crystal structure. The observed transition from a
Kondo insulator to a heavy semimetal, with signatures of linearly dispersing
quasiparticles of low velocity, suggests an interpretation in terms of a
topological Kondo insulator to Weyl-Kondo semimetal \cite{Lai16.1} transition.We
expect these findings to trigger an active search for other such tuning series,
as well as experiments that further probe the newly established Weyl-Kondo
semimetal phase and the nature of its transition to the Kondo insulator phase.

We gratefully acknowledge fruitful discussions with S.\ E.\ Grefe, H.-H.\ Lai,
and E.\ Morosan, helpful advice from K.\ Hradil, W.\ Artner, and D.\ Joshi, and
financial support from the Austrian Science Fund (doctoral program W1243 and
I2535-N27) and the U.S.\ Army Research Office (ARO Grant No.\ W911NF-14-1-0496)
in Vienna, and the ARO (Grant No.\ W911NF-14-1-0525) and the Robert A.\ Welch
Foundation (Grant No.\ C-1411) at Rice. The X-ray measurements were carried out
at the X-Ray Center, Vienna University of Technology.

\vspace{0.2cm}

\noindent $^{\ast}$Corresponding author: {paschen@ifp.tuwien.ac.at}

\vspace{-0.3cm}


\end{document}